\documentclass{article}
\begin{document}

\renewcommand{\a}[0]{\alpha}
\renewcommand{\b}[0]{\beta}
\newcommand{\g}[0]{\gamma}
\newcommand{\G}[0]{\Gamma}
\renewcommand{\d}[0]{\delta}
\newcommand{\e}[0]{\epsilon}
\renewcommand{\k}[0]{\kappa}
\newcommand{\m}[0]{\mu}
\newcommand{\n}[0]{\nu}
\newcommand{\p}[0]{\phi}
\renewcommand{\r}[0]{\rho}
\newcommand{\s}[0]{\sigma}
\renewcommand{\S}[0]{\Sigma}
\renewcommand{\t}[0]{\theta}
\newcommand{\x}[0]{\xi}
\newcommand{\lrb}[0]{\left (}
\newcommand{\rrb}[0]{\right )}

\newcommand{\lsb}[0]{\left [}
\newcommand{\rsb}[0]{\right ]}
\newcommand{\lcb}[0]{\left \{}
\newcommand{\rcb}[0]{\right \}}

\newcommand{\D}[0]{\Delta}
\newcommand{\sa}[0]{\overrightarrow{\s}}
\newcommand{\nn}[0]{\nonumber}
\newcommand{\gph}[0]{(-g)^{\half}}
\newcommand{\gmh}[0]{(-g)^{-\half}}
\newcommand{\tb}[0]{\overline\theta}
\newcommand{\del}[0]{\partial}
\renewcommand{\bar}[0]{\overline}
\newcommand{\be}[0]{\begin{eqnarray}}
\newcommand{\ee}[0]{\end{eqnarray}}

\newcommand{\half}[0]{\frac{1}{2}}
\newcommand{\third}[0]{\frac{1}{3}}
\newcommand{\sixth}[0]{\frac{1}{6}}
\newcommand{\delkap}[0]{\delta_{\kappa}}
\newcommand{\delep}[0]{\delta_{\epsilon}}
\newcommand{\egtl}[1]{\overline{\epsilon}\Gamma_{#1}\theta}
\newcommand{\egtu}[1]{\overline{\epsilon}\Gamma^{#1}\theta}
\newcommand{\xgtl}[1]{\overline{\xi}\Gamma_{#1}\theta}
\newcommand{\xgtu}[1]{\overline{\xi}\Gamma^{#1}\theta}
\newcommand{\dtgtl}[1]{d\overline{\theta}\Gamma_{#1}\theta}
\newcommand{\dtgtu}[1]{d\overline{\theta}\Gamma^{#1}\theta}
\newcommand{\tgdtl}[1]{\overline{\theta}\Gamma_{#1}d\theta}
\newcommand{\tgdtu}[1]{\overline{\theta}\Gamma^{#1}d\theta}
\newcommand{\tgdotl}[1]{\overline{\theta}\Gamma_{#1}\del_{1}\theta}
\newcommand{\tgdotu}[1]{\overline{\theta}\Gamma^{#1}\del_{1}\theta}
\newcommand{\gxl}[2]{(\Gamma_{#1} \xi)_{#2}}
 \newcommand{\gxu}[2]{(\Gamma^{#1} \xi)_{#2}}
\newcommand{\gel}[2]{(\Gamma_{#1} \epsilon)_{#2}}
\newcommand{\geu}[2]{(\Gamma^{#1} \epsilon)_{#2}}
\newcommand{\gtl}[2]{(\Gamma_{#1} \theta)_{#2}}
\newcommand{\gtu}[2]{(\Gamma^{#1} \theta)_{#2}}
\newcommand{\gdotl}[2]{(\Gamma_{#1} \del_{1}\theta)_{#2}}
\newcommand{\gdotu}[2]{(\Gamma^{#1} \del_{1}\theta)_{#2}}

\begin{titlepage}
\thispagestyle{empty}

\begin{flushright}
\end{flushright}
\vspace{5mm}

\begin{center}
{\Large \bf
D-branes and geometrical kappa-symmetry}
\end{center}

\begin{center}
{\large D. T. Reimers}\\
\vspace{2mm}

\footnotesize{
{\it School of Physics, The University of Western Australia\\
Crawley, W.A. 6009, Australia}
} \\
{\tt  reimers@physics.uwa.edu.au}\\
\end{center}
\vspace{5mm}

\begin{abstract}
\baselineskip=14pt

New extended superspaces associated with topological charge algebras of the D-brane are used to construct D-brane actions without worldvolume gauge fields. The actions are shown to be kappa-symmetric under an appropriately chosen right group action.

\end{abstract}

\vfill
\end{titlepage}

\section{Introduction}

Extended superspace action formalisms have been developed for both $p$-branes and D-branes. In the case of $p$-branes, these are based on extended superalgebras which allow the construction of manifestly super-Poincar\'{e} invariant WZ (Wess-Zumino) terms \cite{siegel94,bergshoeff95}. This is related to the classification of cocycles according to their CE (Chevalley-Eilenberg) cohomology \cite{azc89-2}. In standard superspace, the WZ field strength is a nontrivial CE cocycle. However, in appropriately extended superspaces, the same field strength is CE trivial. These are the superspaces which admit the construction of a manifestly super-Poincar\'{e} invariant WZ term. One can similarly construct actions for D-branes in extended superspaces \cite{sak98,sak99,chrys99,sak00}. The basic idea is the same, except that now there are two cocycles. Trivialization of the cocycle associated with the WZ term allows manifestly super-Poincar\'{e} invariant actions to be constructed. Trivialization of the cocycle associated with the NS-NS (Neveu-Schwarz) potential allows the construction of actions without BI (Born-Infeld) worldvolume gauge fields.\\

The Noether charge algebra for $p$-branes is a modification of the standard background superalgebra due to ``quasi-invariance" of the WZ term \cite{azc89}. A similar modification also occurs to the D-brane Noether charge algebra \cite{sorokin97,hammer97}. These algebras are conventionally evaluated by retaining only the bosonic topological charges. Recently, it was noted that if fermionic topological charges are formally retained and used to close the algebra, one recovers extended algebras which allow trivialization of the cocycle \cite{reimers05,reimers05-2,reimers05-D-brane_cocycles}. Furthermore, due to a gauge freedom in the Noether charge algebra, one recovers a ``spectrum" of such algebras \cite{reimers05,reimers05-D-brane_cocycles}. This spectrum contains the known extended algebras as special cases.\\

A defining feature of both $p$-branes and D-branes is the presence of the local $\k$-symmetry. This symmetry allows half the standard fermionic degrees of freedom to be gauged away, and is associated with fermionic supercovariant derivatives \cite{azc91}. As a result, $\k$-symmetry is naturally implemented via a right action of the background supertranslation group \cite{mcarthur00}. The standard superspace action for D-branes contains the BI worldvolume gauge field. The transformation properties of this field under the left group action are postulated to compensate for those of the NS-NS potential. In the extended superspace formalism, this transformation property is explained geometrically by replacing the BI gauge field with a form on the extended superspace with the correct transformation properties. The question is, can the required $\k$ transformation properties of the BI gauge field be similarly explained with group geometry? A clue is that manifest $\k$-symmetry of an extended superspace $p$-brane results from an appropriately chosen right group action \cite{azc04}.\\

In this paper, we derive a set of sufficient conditions for establishing $\k$-symmetry of extended superspace D-brane actions via a right action. The spectrum of D-brane topological charge algebras found in \cite{reimers05-D-brane_cocycles} is then investigated. Subalgebras associated with the NS-NS potential are shown to allow the construction of extended superspace actions for D-branes. $\k$-symmetry of the actions is then established by solving for the required right action.\\

The structure of this paper is as follows. In section \ref{sec:D-branes}, D-brane actions in standard, flat backgrounds are reviewed. In section \ref{sec:D-branes in extended superspaces}, the extended superspace formalism for D-brane actions is reviewed. A set of sufficient conditions for $\k$-symmetry of extended superspace actions is derived. In section \ref{sec:D-branes in extended type IIB superspaces}, the topological charge algebras associated with the NS-NS potential of type IIB D-branes are investigated. It is shown that the associated extended, type IIB superspaces allow the construction of D-brane actions without worldvolume gauge fields. These actions are shown to be $\k$ symmetric by solving for the right action. In section \ref{sec:D-branes in extended type IIA superspaces}, the process is repeated for D-branes in extended, type IIA superspaces. In section \ref{sec:Comments}, some comments on action formalisms are made.

\section{D-branes}
\label{sec:D-branes}

\subsection{Standard actions}

The conventions used in this paper are described fully in \cite{reimers05-D-brane_cocycles}. We work with the standard, flat, background superspaces in d=10. For type IIB superspace it will be assumed that spinor indices are accompanied by an index $I=(1,2)$ which identifies the spinor. The Pauli matrices $(\s_i)_{IJ}$ act upon these indices. $\G^a{}_{\a\b}$ is assumed to be symmetric. The right acting convention for the de Rham differential will be used, and wedge product multiplication of forms is understood.\\

The superalgebra of the supertranslation group is:
\be
    \{Q_{\alpha},Q_{\beta}\}&=&\Gamma^{a}{}_{\alpha\beta}P_{a}.
\ee
The corresponding group manifold can be parameterized:
\be
    \label{2:parametrization}
    g(Z)&=&e^{x^{a}P_{a}}e^{\theta^{\a}Q_{\a}}\\
    Z^{A}&=&\{x^{a},\theta^{\a}\}.\nn
\ee
The left vielbein is defined by:
\be
	L(Z)&=&g^{-1}(Z)dg(Z)\\
	&=&dZ^{M}L_{M}{}^{A}(Z)T_{A}\nn,
\ee
where $T_A$ represents the full set of superalgebra generators. The right vielbein is defined similarly:
\be
	R(Z)&=&dg(Z)g^{-1}(Z)\\
	&=&dZ^{M}R_{M}{}^{A}(Z)T_{A}\nn.
\ee
The left action of the supertranslation group on itself is defined by:
\be
	g(Z')&=&g(\e)g(Z).
\ee
This action is generated by operators $Q_A$ (``left generators"). One finds:
\be
	\label{left generators for Z}
	\d Z^M&=&\e^AQ_{A}Z^M\\
	&=&\e^AR_{A}{}^{M}\nn,
\ee
where $R_{A}{}^{M}$ are the inverse right vielbein components, defined by:
\be
	R_{A}{}^{M}R_{M}{}^{B}=\d_{A}{}^{B}.
\ee
Forms that are invariant under a global left action will be called ``left invariant." The left vielbein components are left invariant by construction.\\

The right group action is defined by:
\be
	g(Z')&=&g(Z)g(\e).
\ee
The corresponding superspace transformation is generated by operators $D_A$. One finds:
\be
	\label{right generators for Z}
	\d Z^M&=&\e^AD_{A}Z^M\\
	&=&\e^AL_{A}{}^{M}\nn,
\ee
where $L_{A}{}^{M}$ are the inverse left vielbein components, defined by:
\be
	L_{A}{}^{M}L_{M}{}^{B}=\d_{A}{}^{B}.
\ee
$D_A$ are commonly known as ``supercovariant derivatives" since they commute with the $Q_A$. Unlike the $Q_{A}$, they do not generate global symmetries of the action. However, they do play a role in the local $\k$-symmetry.\\

D$p$-branes are $\k$-symmetric, $p+1$ dimensional ``worldvolumes" embedded in the background superspace. D$p$-branes in type IIA superspace exist only for $p$ even, while those in type IIB superspace exist only for $p$ odd. Actions for D-branes have been developed in both flat and more general backgrounds \cite{Schmidhuber96_D-brane_actions,aganagic97,cederwall96,Bergshoeff96}. We now present the action with the conventions adopted in this paper.\\

Let the worldvolume be parameterized by coordinates $\s^i$. The worldvolume metric $g_{ij}$ is defined using the pullbacks of the left vielbein components:
\be
    L_{i}{}^{A}&=&\del_{i}Z^{M}L_{M}{}^{A}\\
    g_{ij}&=&L_{i}}{^{a}L_{j}{}^{b}\eta_{ab}\nn.
\ee
The action consists of two terms:
\be
	\label{standard action}
	S&=&S_{DBI}+S_{WZ}.
\ee
The DBI term is:
\be
	\label{non manifest action}
	S_{DBI}=-\int d^{p+1}\s\sqrt{-\mathsf{det}(g_{ij}+F_{ij})}.
\ee
$F$ is a $2$-form:
\be
	F&=&B-dA.
\ee
$A=d\s^iA_i$ is the BI worldvolume gauge field, which is a $1$-form defined only on the worldvolume. The NS-NS potential $B$ is a superspace $2$-form defined by:
\be
	dB&=&H,
\ee
where $H$ is the left invariant NS-NS $3$-form field strength. For type IIA superspace, $H$ is:
\be
	H&=&\half L^a d\bar\t\G_{11}\G_ad\t,
\ee
while for type IIB:
\be
	H&=&-\half L^a d\bar\t\G_a \s_3 d\t.
\ee
For type IIA superspace, closure of $H$ requires the ``standard" Fierz identity \cite{aganagic97}:
\be
	\G^a{}_{(\a\b}(\G_{11}\G_a)_{\g\d)}&=&0,
\ee
while for type IIB:
\be
	\G^a{}_{(\a\b}(\G_a\s_3)_{\g\d)}&=&0.
\ee

The second term in the action is the WZ term:
\be
	S_{WZ}&=&\int b.
\ee
It is defined by the formal sum of forms:
\be
	b=\breve b e^{F}.
\ee
The form of degree $p+1$ is selected from this sum and the integral is then performed over the worldvolume of the brane. We denote the form of a specific degree in a formal sum by a number in brackets. For example:
\be
	\breve b&=&\oplus \breve b^{(n)}.
\ee
The R-R (Ramond) potentials $\breve b^{(n)}$ are defined by:
\be
	\label{R-R potential defn}
	R=d\breve b +\breve b H.
\ee
The R-R field strengths $R^{(n)}$ are left invariant superspace forms:
\be
	R^{(n)}=(-1)^pd\bar\t S^{(n-2)}d\t,
\ee
where for type IIA superspace the $S^{(n)}$ are given by:
\be
	S^{(n)}&=&\frac{1}{2n!}L^{a_1}\ldots L^{a_n}\G_{a_1 \ldots a_n}\G_{11}{}^{[\frac{n}{2}+1]},
\ee
while for type IIB:
\be
	S^{(n)}&=&\frac{1}{2n!}L^{a_1}\ldots L^{a_n}\G_{a_1 \ldots a_n}\s_3{}^{[\frac{n+1}{2}+1]}\s_1.
\ee
From (\ref{R-R potential defn}) it follows that the total field strength for the WZ term is the degree $p+2$ piece of:
\be
	h&=&db\\
	&=&Re^F\nn.
\ee
Closure of $h$ is equivalent to some more general Fierz identities. For type IIA superspace these are:
\be
	(m-1)(\G_{11}{}^{\frac{m}{2}}\G_{[a_1\ldots a_{m-2}})_{(\a\b}(\G_{11}\G_{a_{m-1}]})_{\g\d)}&&\\
		-\G^{a_m}{}_{(\a\b}(\G_{11}{}^{\frac{m+2}{2}}\G_{a_1\ldots a_m})_{\g\d)}&=&0,\nn
\ee
while for type IIB:
\be
	(m-1)(\G_{[a_1\ldots a_{m-2}}\s_3{}^{\frac{m+1}{2}}\s_1)_{(\a\b}(\G_{a_{m-1}]}\s_3)_{\g\d)}&&\\
		+\G^{a_m}{}_{(\a\b}(\G_{a_1\ldots a_m}\s_3{}^{\frac{m+3}{2}}\s_1)_{\g\d)}&=&0.\nn
\ee
Most of these can be shown to hold by repeated use of the $m=2$ identity \cite{aganagic97,cederwall96}.\\

It is a somewhat mysterious feature of the standard superspace D-brane action that the BI gauge field also transforms under the left action of the supertranslation group. This transformation is determined by the requirement that the potential $F$ must be left invariant. Since $[d,Q_A]=0$, we must therefore require:
\be
	dQ_A A=Q_A B.
\ee
From the left invariance of $H$ it follows that:
\be
	\label{Q_A B=dW_A}
	Q_A B&=&-dW_A
\ee
for some set of $1$-forms $W_A$. Hence:
\be
	\label{Q_A A=-W_A}
	Q_A A_i=-(W_A)_i
\ee
is the required transformation of the BI gauge field. Furthermore, since $H$ is CE nontrivial, there does not exist a potential $B$ such that $Q_A B=0$ for all $Q_A$ \cite{chrys99,sak00}.\\

D-branes also have a local $\k$-symmetry. Let $V_\k$ denote the infinitesimal vector field generated by a $\k$ transformation. This field is assumed to take the form:
\be
	\label{kappa transformation std}
	V_\k{}^M&=&\e^\a L_\a{}^M\\
	\bar\e_\a&=&(\bar\k P_-)_\a\nn,
\ee
where $\k^\a(\s)$ are local worldvolume parameters and $P_-$ is a projection operator to be determined. $\k$-symmetry of the action requires that:
\be
	\label{delta kappa F}
	\d_\k F=i_{V_\k}H,
\ee
where $i$ is the interior derivation. The $\k$ variation of a superspace form $Y$ is identically:
\be
	\d_\k Y&=&i_{V_\k} dY+di_{V_\k} Y.
\ee
Equation (\ref{delta kappa F}) then implies that the $\k$ transformation of the BI gauge field is:
\be
	\label{delta kappa A}
	\d_\k A_i&=&(i_{V_\k}B)_i
\ee
up to a total derivative. Using (\ref{delta kappa F}) one finds:
\be
	\label{delta kappa b}
	\d_\k b=i_{V_\k}Re^F+d(i_{V_\k}\breve b e^F).
\ee

We now summarize the proof of $\k$-symmetry given in \cite{aganagic97} using the conventions adopted in this paper. For type IIA superspace, define the matrix valued forms:
\be
	\rho&=&e^F \sum_{n\ \mathsf{odd}}S^{(n)}\\
	T&=&e^F \sum_{n\ \mathsf{even}}S^{(n)}\nn,
\ee
while for type IIB:
\be
	\rho&=&e^F \sum_{n\ \mathsf{even}}S^{(n)}\\
	T&=&e^F \sum_{n\ \mathsf{odd}}S^{(n)}\nn.
\ee
$T$ is related to the WZ field strength via:
\be
	h&=&(-1)^p d\bar\t T d\t.
\ee
Then define the matrix valued densities:
\be
	\label{matrix valued densities}
	\tilde \rho&=&\frac{1}{(p+1)!}\tilde\e^{i_{p+1}\ldots i_1}\rho_{i_1\ldots i_{p+1}}\\
	T^i&=&\frac{1}{p!}\tilde\e^{i_p\ldots i_1i}T_{i_1\ldots i_p}\nn,
\ee
where $\tilde \e$ is the antisymmetric Levi-Civita symbol. The associated matrix:
\be
	\g=-\frac{\tilde \rho}{\mathcal{L}_{DBI}}
\ee
can be shown to be idempotent \cite{aganagic97}:
\be
	\g^2=1.
\ee
The $\k$ variation of the Lagrangian is then found to be:
\be
	\label{delta kappa Lagrangian}
	\d_\k \mathcal{L}=-\bar\e(1+\g)T^i \del_i\t.
\ee
Setting:
\be
	P_-=\half(1-\g)
\ee
then gives the required $\k$-symmetry. Since $\g$ is traceless, the $\pm 1$ eigenvalues are present in equal numbers. This allows half the $\t^\a$ to be gauged away using the local parameter in (\ref{kappa transformation std}).

\section{D-brane actions on extended superspaces}
\label{sec:D-branes in extended superspaces}

Closed forms can be classified using CE group cohomology \cite{azc89-2}. A closed form $Y=dX$ on the background superspace is CE trivial if there exists a potential $X$ such that
\be
	Q_AX&=&0.
\ee
Otherwise $Y$ is CE nontrivial. CE nontrivial forms will be termed ``cocycles," and CE trivial forms ``coboundaries."\\

Note that even for standard D-brane actions one has $H=dF$, and:
\be
	Q_AF&=&0.
\ee
Since $F$ is not a form on the background superspace, the CE cohomology does not apply. However, this preempts the philosophy for D-brane actions on extended superspaces. One wishes to find an extended superspace which allows the construction of a left invariant potential for $H$ \cite{sak98,sak99,chrys99}:
\be
	\label{trivial F def}
	dF'&=&H\\
	Q_AF'&=&0.\nn
\ee
Since $F'$ is now a form on the extended background superspace, this represents a ``trivialization of the cocycle". Denote the standard D-brane action (\ref{standard action}) as $S[x,\t,F]$. The extended superspace D-brane action is then simply:
\be
	\label{extended action}
	S'[x,\t,\ldots]&=&S[x,\t,F'],
\ee
where $\ldots$ indicates extra superspace coordinates. In this way, the BI gauge field is replaced by a form involving extra superspace coordinates. This demonstrates one of the strengths of the extended superspace formalism: it describes the left group transformation property (\ref{Q_A A=-W_A}) of the BI gauge field \textit{geometrically}. One of the defining properties of the D-brane is the required existence of local $\k$-symmetry. In the same spirit as for the left group transformations, we now seek a geometrical description of the $\k$ transformations of the BI gauge field.\\

It is a general property of $\k$-symmetry that it is generated by a right action of the background supertranslation group \cite{mcarthur00}. Let us first consider the standard D-brane action. In this case the $\k$-symmetry (\ref{kappa transformation std}) is generated by the right action:
\be
	\label{kappa properties std}
	g(Z+V_\k{}^M)&=&g(Z)g(\e)\\
	\e^A&=&\{0,\e^\a\},\nn
\ee
where $\e^\a$ is defined in (\ref{kappa transformation std}). This exhibits another property of $\k$ transformations: the parameter $\e^a$ of the right action (corresponding to the superalgebra generator $P_a$) vanishes. These properties go hand in hand with the structure of the field strengths $H$ and $h$. From (\ref{delta kappa F}) and (\ref{delta kappa b}), the $\k$ variations of $F$ and $b$ are determined by the explicit structure of $i_{V_K}H$ and $i_{V_K}R^{(n)}$ respectively. Both $H$ and $R^{(n)}$ have the structure:
\be
	Y&=&L^{a_1}\ldots L^{a_m}d\bar\t\G_{a_1\ldots a_m}Sd\t,
\ee
where $S$ is either $\G_{11}$ or a Pauli matrix. This yields the interior derivations:
\be
	\label{i V H and i V R}
	i_{V_\k}Y&=&-2L^{a_1}\ldots L^{a_m}\bar\e \G_{a_1\ldots a_m}Sd\t
\ee
\textit{provided} that $V_\k$ results from a right action with the parameters (\ref{kappa properties std}). The variation $\d_\k g_{ij}$ of the metric does not contain derivatives of $\e^\a$. The resulting structure of $\d_\k F_{ij}$ is then such that it combines nicely with $\d_\k g_{ij}$.\\

For the extended superspace formalism, it is first noted that manifest $\k$-symmetry of an extended superspace $p$-brane Lagrangian can be achieved with a right action \cite{azc04}. We will use a similar mechanism to establish $\k$-symmetry for extended superspace D-brane actions. In the case of $p$-branes, the extra superspace coordinates appear in the action only through a total derivative (except in the scale invariant approach of \cite{azc04}). The mechanism is then optional since the variation of the Lagrangian will be a total derivative irrespective of the way $\k$ transformations of extra coordinates are defined. However, in the case of D-branes the mechanism \textit{must} be used since extra coordinates appear in the DBI term.\\

Let $Z^{M}$ now denote both the standard and extra coordinates of the extended superspace:
\be
	Z^{M}&=&\{Z^{\widetilde M},Z^{\check M}\}\\
	&=&\{x^m,\t^\m,Z^{\check M}\}\nn.
\ee
We first note that topological charge algebras are extensions of the standard superalgebra by an ideal, and that the algebras known to allow construction of extended superspace actions appear in the spectrum of topological charge algebras \cite{reimers05,reimers05-2,reimers05-D-brane_cocycles}. We therefore assume that the background superalgebra is an extension of the standard background superalgebra by an ideal. The following properties follow:
\begin{itemize}
\item
The standard coordinate blocks $L_{\widetilde M}{}^{\widetilde A}$ and $L_{\widetilde A}{}^{\widetilde M}$ retain their original structure.
\item
$L_{\check A}{}^{\widetilde M}=0$.
\end{itemize}
This leads us to seek a new infinitesimal vector field $V'_\k$ generated by the right action:
\be
	\label{kappa symmetry properties extended}
	V'_\k{}^{M}&=&\e^{A}L_{A}{}^{M}\\
	\e^{A}&=&\{0,\e^\a,\e^{\check A} \},\nn
\ee
where $\e^\a$ is the same as in (\ref{kappa transformation std}), and $\e^{\check A}$ are to be determined. Firstly, the $\k$ variation of the standard coordinates is then unchanged. Thus:
\be
	\d'_\k g_{ij}&=&\d_\k g_{ij}.
\ee
Secondly, although the proof of $\k$-symmetry for the standard D-brane depends on the explicit structure of $\d_\k F$, it does not depend on that of $F$ itself (see, for example \cite{aganagic97}). This allows us to use the $\k$-symmetry mechanism of the standard action in proving $\k$-symmetry of the extended action. If the parameters $\e^{A}$ of the right action are chosen such that:
\be
	\label{kappa symmetry condition}
	di_{V'_\k}F'=0,
\ee
then the transformation $\d'_\k F'=i_{V'_\k}H$ is obtained. Since $H$ and $R^{(n)}$ are constructed only from $d\t$ and $L^a$, the extra parameters in (\ref{kappa symmetry properties extended}) do not affect the relevant interior derivations. We thus have $i_{V'_\k}H=i_{V_\k}H$, and therefore $\d'_\k F'=\d_\k F$. We also have $i_{V'_\k}R^{(n)}=i_{V_\k}R^{(n)}$, from which it follows that the variation $\d'_\k b'$ of the new WZ form is equal to the standard variation $\d_\k b$ up to a total derivative. This derivative is ignored in the same way as the second term of (\ref{delta kappa b}). We then have $\d'_\k \mathcal{L'}$ equal to $\d_\k \mathcal{L}$ from (\ref{delta kappa Lagrangian}) up to a total derivative. Equations (\ref{trivial F def}), (\ref{kappa symmetry properties extended}) and (\ref{kappa symmetry condition}) are therefore sufficient conditions to establish $\k$-symmetry of the extended superspace D-brane actions.

\section{D-branes in extended type IIB superspaces}
\label{sec:D-branes in extended type IIB superspaces}

\subsection{Cocycle trivialization}

We wish to find extended superspaces which admit left invariant solutions of the equation:
\be
	dF'&=&H\\
	&=&-\half L^a d\bar\t\G_a \s_3 d\t.\nn
\ee
These extended type IIB superspaces will allow the construction of D-brane actions without BI gauge fields via the action (\ref{extended action}). We consider the spectrum of topological charge algebras of the D-strings derived in \cite{reimers05-D-brane_cocycles}. The subset of this spectrum associated with the NS-NS potential of type IIB D-branes is\footnote{The free phase angle in the algebra of \cite{reimers05-D-brane_cocycles} should be set equal to that of the $SO(2)$ frame of the type IIB action being used, which in this case is zero. This results in a minimal set of extra generators. Differences in sign amount to the rescaling $(Q_\a,P_a)\rightarrow (-Q_\a,-P_a)$.}:
\be
	\label{IIB NS-NS algebra spectrum}
	\big \{Q_\a,Q_\b\big \}&=&\G^a{}_{\a\b}P_a+\bigg [E-\half\bigg ](\G_a\s_3)_{\a\b}\S^a\\
		&&-\bigg [E-\frac{1}{4}\bigg ]\S_{\a\b}\nn\\
	\big [Q_\a,P_b\big ]&=&-E(\G_b\s_3)_{\a\b}\S^\b\nn\\
	\big [Q_\a,\S^b\big ]&=&\G^b{}_{\a\b}\S^\b\nn\\
	\big [Q_\a,\S_{\b\g}\big ]&=&-\Big [\G^a{}_{\a(\b}(\G_a\s_3)_{\g)\d}-\G^a{}_{\d(\b}(\G_a\s_3)_{\g)\a}\Big ]\S^\d\nn,
\ee
where $E$ is a free constant. The generators will be associated with the following left vielbein components:
\be
\{P_a,Q_\a,\S^a,\S^\a,\S_{\a\b}\}&\rightarrow &\{L^a,L^\a,L_a,L_\a,L^{\a\b}\}.
\ee
The Maurer-Cartan equations following from (\ref{IIB NS-NS algebra spectrum}) are:
\be
	dL^\a&=&d\t^\a=0\\
	dL^a&=&-\half d\bar\t\G^ad\t\nn\\
	dL_a&=&-\half\bigg [E-\half\bigg ] d\bar\t\G_a\s_3d\t\nn\\
	dL_\a&=&EL^b(\G_b\s_3d\t)_\a-L_b(\G^bd\t)_\a\nn\\
		&&+L^{\b\g}\Big [(\G^bd\t)_{\b}(\G_b\s_3)_{\g\a}-\G^b{}_{\a\b}(\G_b\s_3d\t)_{\g}\Big ]\nn\\
	dL^{\a\b}&=&\half\bigg [E-\frac{1}{4}\bigg ] d\t^\a d\t^\b\nn.
\ee
The required potential $F'$ on the extended superspace can be found by forming all possible super-Poincar\'{e} invariant $2$-forms of dimension two, and then equating coefficients in the equation $dF'=H$. Since the vielbein components are already left invariant, it suffices to consider all possible Lorentz invariant bilinear combinations. It is useful to introduce an arbitrary constant $K$, and then find the general solution of the equation:
\be
	dF'&=&-KL^ad\bar\t\G_a\s_3d\t.
\ee
The action being used then corresponds to $K=\half$. The solution is found to exist for all $E\neq\frac{1}{4}$, and contains a free parameter $C$:
\be
	\label{IIB explicit solution for F}
	F'&=&C_1L^aL_a+C_2L^\a L_\a+C_3L^aL^{\a\b}(\G_a\s_3)_{\a\b}+C_4L_aL^{\a\b}\G^a{}_{\a\b}\\
		&&+C_5L^{\a\b}\G^a{}_{\a\b}L^{\g\d}(\G_a\s_3)_{\g\d}\nn\\
	C_1&=&-4K-\frac{C}{4}(4E-1)\nn\\
	C_2&=&2K\nn\\
	C_3&=&-\frac{C}{2}(2E-1)\nn\\
	C_4&=&C\nn\\
	C_5&=&-\frac{2C(2E-1)}{4E-1}\nn.
\ee
The presence of the free parameter means that the solution has the form:
\be
	F'=-4KL^aL_a+2KL^\a L_\a+CF'_{\mathsf{homog}},
\ee
where $dF'_{\mathsf{homog}}=0$. This type of solution is familiar for differential equations. The first two terms are the unique particular solution, while the last is the solution of the associated homogenous equation. $F'_{\mathsf{homog}}$ is a Lorentz invariant $2$-cocycle. It is explicitly obtained by setting $K=0$ in the solution (\ref{IIB explicit solution for F}) for $F'$.\\

There are two special cases of the algebra (\ref{IIB NS-NS algebra spectrum}) where generators become redundant. The first is for $E=\frac{1}{4}$ (the singular case of the general solution). In this case $\S^{\a\b}$ appears nowhere on the RHS of a bracket and may therefore be excluded from the algebra entirely. Upon rescaling, this algebra is equivalent to ones considered in \cite{sak98,sak99,sak00}. The unique solution for $F'$ is then:
\be
	\label{IIB two term solution for F}
	F'&=&-4KL^aL_a+2KL^\a L_\a.
\ee
The second special case is for $E=\half$, which allows $\S^a$ to be excluded from the algebra. The algebra in this case yields the unique solution:
\be
	\label{IIB one term solution for F}
	F'&=&2KL^\a L_\a.
\ee

Therefore, topological charge algebras associated with the NS-NS potential of the standard, type IIB action can be used to construct D-brane actions on extended, type IIB superspaces. In the general case where all generators are present, the existence of a Lorentz invariant $2$-cocycle of dimension two indicates that the superspace is extended more than is necessary for cocycle trivialization. The algebras in the two special cases $E=\{\frac{1}{4},\frac{1}{2}\}$ may be viewed as ``minimal extensions" in this regard.

\subsection{$\k$-symmetry}

To show that the new $F'$ potentials yield D-brane actions (\ref{extended action}) which are $\k$-symmetric we need to find a right action satisfying the conditions (\ref{kappa symmetry properties extended}) and (\ref{kappa symmetry condition}). Let us first consider the general case where all the generators are present. Associate the parameters $\e^A$ of the right action with the generators of the algebra (\ref{IIB NS-NS algebra spectrum}) as:
\be
	\{P_a,Q_\a,\S^a,\S^\a,\S_{\a\b}\}&\rightarrow &\{\e^a,\e^\a,\e_a,\e_\a,\e^{\a\b}\}.
\ee
From (\ref{kappa symmetry properties extended}), we already know that $\e^a=0$, while $\e^\a$ is given by (\ref{kappa transformation std}). To find the remaining parameters one needs to use properties of the matrices $\G^a$ and $\s_i$ under trace\footnote{Since we are working with compound IIB indices, $\G^\a{}_\b$ is actually shorthand for $\G^{\a I}{}_{\b J}=\G^\a{}_\b\d^I{}_J$. This yields twice the ordinary trace of gamma matrices.} in d=10:
\be
	(\G_a\s_3\G_b\s_3)^\a{}_\a&=&64\eta_{ab}\\
	(\G_a\s_3\G_b)^\a{}_\a&=&0\nn.
\ee
By requiring $di_{V_\k}F'=0$ one then finds a solution for the parameters $\e^A=\{0,\e^\a,\e_a,0,\e^{\a\b}\}$, with:
\be
	\e_a&=&\lrb\frac{C_2}{C_1+\frac{C_3C_4}{C_5}}\rrb \eta_{ab}L_i{}^bg^{ij}\e^\a L_{j\a}\\
	\e^{\a\b}&=&-\lrb \frac{C_4}{64C_5}\rrb b_a(\G^a\s_3)^{\a\b}\nn.
\ee

In the case $E=\frac{1}{4}$ (with $\S^{\a\b}$ absent), the solution for $F'$ is given by (\ref{IIB two term solution for F}). One finds the solution for the parameters $\e^A=\{0,\e^\a,\e_a,0\}$, with:
\be
	\e_a&=&\lrb \frac{C_2}{C_1}\rrb\eta_{ab}L_i{}^bg^{ij}\e^\a L_{j\a}.
\ee
$\k$-symmetry in this case is analogous to manifest $\k$-symmetry of the GS superstring \cite{azc04}.\\

The mechanism used in the previous two cases relies on the existence of a matrix $K_a{}^i$ such that:
\be
	L_i{}^aK_a{}^j=\d_i{}^j.
\ee
Indeed, the results were obtained by setting:
\be
	K_a{}^j=\eta_{ab}L_k{}^bg^{kj}.
\ee
In the case $E=\frac{1}{2}$ (with $\S^a$ absent), the solution for $F'$ is given by (\ref{IIB one term solution for F}). One finds that the same mechanism now relies on the existence of a matrix $K_\a{}^i$ such that:
\be
	L_i{}^\a K_\a{}^j=\d_i{}^j.
\ee
One can consider, for example:
\be
	\label{fermionic metrics}
	K_\a{}^j&=&C_{\a\b}\del_k\t^\b\tilde g^{kj}\\
	\tilde g_{ij}&=&\del_i\bar\t\del_j\t.
\ee
Whilst the nondegeneracy of $g_{ij}$ holds for all ``non-lightlike" solutions of the equations of motion, nondegeneracy of $\tilde g_{ij}$ holds only for solutions with nonvanishing $\t^a$. This is a severe restriction (which for example excludes all solutions of the bosonic action). It thus appears that the standard right action mechanism for $\k$-symmetry does not apply when $\S^a$ is absent from the algebra.\\

Of the three solutions (\ref{IIB explicit solution for F}), (\ref{IIB two term solution for F}) and (\ref{IIB one term solution for F}) for $F'$, the last two make use of algebras which are minimal extensions of the background, while the first two admit $\k$-symmetry via the standard mechanism. Using the extended background superalgebra with $E=\frac{1}{4}$ and $\S^{\a\b}$ absent (with the solution (\ref{IIB two term solution for F}) for $F'$) therefore appears to be a good choice for extended superspace D-brane actions.

\section{D-branes in extended type IIA superspaces}
\label{sec:D-branes in extended type IIA superspaces}

\subsection{Cocycle trivialization}

We wish to find extended type IIA superspaces which admit left invariant solutions of the equation:
\be
	dF'&=&KL^ad\bar\t\G_{11}\G_ad\t.
\ee
The action being used corresponds to $K=\half$. These superspaces will then allow D-brane actions without BI gauge fields to be defined via the action (\ref{extended action}). We consider the spectrum of topological charge algebras associated with the NS-NS potential of the D-membrane, derived in \cite{reimers05-D-brane_cocycles}. This spectrum is:
\be
	\label{IIA algebra spectrum}
	\big \{Q_\a,Q_\b\big \}&=&\G^a{}_{\a\b}P_a-\Big [\bigg [E_2-\half\bigg ](\G_{11}\G_a)_{\a\b}+E_1\G_{a\a\b}\Big ]\S^a\\
		&&+\bigg [E_2-\frac{1}{4}\bigg ]\S_{\a\b}\nn\\
	\big [Q_\a,P_b\big ]&=&\Big [E_1\G_{b\a\b}+E_2(\G_{11}\G_b)_{\a\b}\Big ]\S^\b\nn\\
	\big [Q_\a,\S^b\big ]&=&\G^b{}_{\a\b}\S^\b\nn\\
	\big [Q_\a,\S_{\b\g}\big ]&=&-\Big [\G^a{}_{\a(\b}(\G_{11}\G_a)_{\g)\d}-\G^a{}_{\d(\b}(\G_{11}\G_a)_{\g)\a}\Big ]\S^\d\nn,
\ee
where $E_1$ and $E_2$ are free constants. Again associate the generators with the left vielbein components:
\be
\{P_a,Q_\a,\S^a,\S^\a,\S_{\a\b}\}&\rightarrow &\{L^a,L^\a,L_a,L_\a,L^{\a\b}\}.
\ee
The resulting Maurer-Cartan equations are:
\be
	dL^\a&=&0\\
	dL^a&=&-\half d\bar\t\G^ad\t\nn\\
	dL_a&=&\half\bigg [E_2-\half\bigg ] d\bar\t\G_{11}\G_ad\t+\half E_1d\bar\t\G_ad\t\nn\\
	dL_\a&=&-E_1L^b(\G_bd\t)_\a-E_2L^b(\G_{11}\G_bd\t)_\a-L_b(\G^bd\t)_\a\nn\\
		&&+L^{\b\g}\Big [(\G^bd\t)_{\b}(\G_{11}\G_b)_{\g\a}-\G^b{}_{\a\b}(\G_{11}\G_bd\t)_{\g}\Big ]\nn\\
	dL^{\a\b}&=&-\half\bigg [E_2-\frac{1}{4}\bigg ] d\t^\a d\t^\b\nn.
\ee

The calculations proceed in the same way as for the type IIB case. The general solution exists for all $E_2\neq\frac{1}{4}$, and again contains a free parameter $C$:
\be
	\label{IIA general solution for F}
	F'&=&C_1L^aL_a+C_2L^\a L_\a+C_3L^aL^{\a\b}\G_{a\a\b}+C_4L^aL^{\a\b}(\G_{11}\G_a)_{\a\b}\\
		&&+C_5L_aL^{\a\b}\G^a{}_{\a\b}+C_6L^{\a\b}\G^a{}_{\a\b}L^{\g\d}(\G_{11}\G_a)_{\g\d}\nn\\
	C_1&=&-4K+\frac{C}{4}(4E_2-1)\nn\\
	C_2&=&2K\nn\\
	C_3&=&E_1C\nn\\
	C_4&=&\frac{C}{2}(2E_2-1)\nn\\
	C_5&=&C\nn\\
	C_6&=&-\frac{2C(2E_2-1)}{4E_2-1}\nn.
\ee
The free parameter $C$ again shows the presence of a Lorentz invariant $2$-cocycle, of dimension two, on the extended superspace. This cocycle is given by the solution (\ref{IIA general solution for F}) for $F'$ with $K=0$.\\

The first special case of the algebra (\ref{IIA algebra spectrum}) is $E_2=\frac{1}{4}$, which allows $\S^{\a\b}$ to be excluded. In the gauge $E_1=0$, this algebra corresponds to one used in \cite{chrys99}. The unique solution for $F'$ (which exists for \textit{all values} of $E_1$) is found to be:
\be
	F'&=&-4KL^aL_a+2KL^\a L_\a.
\ee
The second special case is when $E_1=0$ and $E_2=\half$, in which case $\S^a$ may be excluded. The unique solution is then:
\be
	F'&=&2KL^\a L_\a.
\ee

Therefore, topological charge algebras associated with the NS-NS potential of the standard, type IIA action can be used to define D-brane actions on their associated superspaces. The two special cases yield extensions of the standard background superalgebra which are minimal (in the sense of cocycle trivialization).

\subsection{$\k$-symmetry}

The calculations for the $\k$-symmetries proceed similarly to the type IIB case. One now needs the trace properties:
\be
	(\G_{11}\G_a\G_{11}\G_b)^\a{}_\a&=&-32\eta_{ab}\\
	(\G_{11}\G_a\G_b)^\a{}_\a&=&0\nn.
\ee
Associate the parameters $\e^A$ of the right action with the generators of the superalgebra as:
\be
	\{P_a,Q_\a,\S^a,\S^\a,\S_{\a\b}\}&\rightarrow &\{\e^a,\e^\a,\e_a,\e_\a,\e^{\a\b}\}.
\ee
In the general case where all the generators are present one finds the solution $\e^A=\{0,\e^\a,\e_a,0,\e^{\a\b}\}$, with:
\be
	\e_a&=&\lrb \frac{C_2}{C_1+\frac{C_4C_5}{C_6}}\rrb \eta_{ab}L_i{}^bg^{ij}\e^\a L_{j\a}\\
	\e^{\a\b}&=&-\lrb \frac{C_5}{32C_6}\rrb b_a(\G_{11}\G^a)^{\a\b}\nn.
\ee
When $\S^{\a\b}$ is absent ($E_2=\frac{1}{4}$), one again obtains $\e^A=\{0,\e^\a,\e_a,0\}$, with:
\be
	\e_a&=&\lrb \frac{C_2}{C_1}\rrb \eta_{ab}L_i{}^bg^{ij}\e^\a L_{j\a}.
\ee
The case where $\S^a$ is absent ($E_1=0$ and $E_2=\half$) again does not admit $\k$-symmetry via the mechanism considered here.\\

Therefore, in the type IIA case there is a \textit{one parameter spectrum} of algebras ($E_1,E_2)=(E_1,\frac{1}{4}$) characterized as ``minimal extensions" that admit $\k$-symmetry via a right action.

\section{Comments}
\label{sec:Comments}

New superalgebras resulting from the topological charge algebras of the standard D-brane action were shown to allow the construction of D-brane actions on extended superspaces. In all but one discrete case, these actions were shown to admit $\k$-symmetry via an appropriately chosen right group action. We observe that there is a correspondence between \textit{manifest} symmetries of Green-Schwarz superstring actions and (non-manifest) symmetries of the D-brane action. In the case of the superstring, one constructs a WZ $2$-form on an extended background superspace. One requires that this form be super-Poincar\'{e} invariant, and also be such that it admits manifest $\k$-symmetry of the action. In type II superspaces, analogous solutions appear which can be used to replace the $F$ field of D-branes. Via the arguments presented in section \ref{sec:D-branes in extended superspaces}, these solutions yield super-Poincar\'{e} invariant, $\k$-symmetric actions without BI gauge fields.\\

Concerning gauge symmetries and the extended superspace formalism, we point out that $\k$-symmetry is not the full story. The ``price" one has paid for an action without worldvolume gauge fields is the introduction of extra superspace degrees of freedom. If the two action formalisms are to be equivalent, they must possess the same number of degrees of freedom. There must then exist gauge symmetries which allow the extra superspace degrees of freedom to be reduced to those of the BI gauge field. This issue has been considered in \cite{azc01}.

\subsection{Acknowledgements}
I would like to thank I. N. McArthur for helpful suggestions and critical reading of the manuscript.

\bibliographystyle{hieeetr}
\bibliography{double_complex}
\end{document}